\documentclass[prd,preprint,showpacs,preprintnumbers,amsmath,amssymb,eqsecnum,nofootinbib,groupedaddress,superscriptaddress]{revtex4-1}
\usepackage{dcolumn}
\usepackage{amsfonts,mathrsfs}
\usepackage{bm}
\usepackage{subfigure}
\usepackage{url}
\usepackage{ulem}
\usepackage[dvipdfmx]{hyperref,graphicx,color}

\begin{document}
\preprint{RUP-22-21}
%
\title{General formulae for the periapsis shift of a quasi-circular orbit 
in static spherically symmetric spacetimes and 
the active gravitational mass density
}
\author{Tomohiro Harada}
\email{harada@rikkyo.ac.jp}
\affiliation{Department of Physics, Rikkyo University, Toshima,
Tokyo 171-8501, Japan}
\author{Takahisa Igata}
\email{takahisa.igata@gakushuin.ac.jp}
\affiliation{Department of Physics, 
Gakushuin University, Mejiro, Toshima, Tokyo 171-8588, Japan}
\author{Hiromi Saida}
\email{saida@daido-it.ac.jp}
\affiliation{Daido University, Nagoya, Aichi 457-8530, Japan}
\author{Yohsuke Takamori}
\email{takamori@wakayama-nct.ac.jp}
\affiliation{National Institute of Technology (KOSEN),
Wakayama College, Gobo, Wakayama 644-0023, Japan}
\date{\today}
\begin{abstract}
We study the periapsis shift of a quasi-circular orbit in general static spherically symmetric spacetimes. We derive two formulae in full order with respect to the gravitational field, one in terms of the gravitational mass $m$ and the Einstein tensor and the other in terms of the orbital angular velocity and the Einstein tensor. These formulae reproduce the well-known ones for the forward shift in the Schwarzschild spacetime. In a general case, the shift deviates from that in the vacuum spacetime due to a particular combination of the components of the Einstein tensor at the radius $r$ of the orbit. The formulae give a backward shift due to the extended-mass effect in Newtonian gravity. In general relativity,  in the weak-field and diffuse regime, the active gravitational mass density, $\rho_{A}=(\epsilon+p_{r}+2p_{t})/c^{2}$, plays an important role, where $\epsilon$, $p_{r}$, and $p_{t}$ are the energy density,  the radial stress, and the tangential stress of the matter field, respectively. We show that the shift is backward if $\rho_{A}$ is beyond a critical value $\rho_{c}\simeq 2.8\times 10^{-15} \mbox{g}/\mbox{cm}^{3} (m/M_{\odot})^{2}(r/\mbox{au})^{-4}$, while a forward shift greater than that in the vacuum spacetime instead implies $\rho_{A}<0$, i.e., the violation of the strong energy condition, and thereby provides evidence for dark energy. We obtain new observational constraints on $\rho_{A}$ in the Solar System and the Galactic Centre.
\end{abstract}
\maketitle
\tableofcontents

\newpage

\section{Introduction \label{sec:introduction}}

The perihelion shift of 
Mercury 
is one of the first classical predictions 
of general relativity by Einstein (1916)~\cite{Einstein:1916vd}, which has been 
the most successful theory of gravity so far. See also Ref.~\cite{Will:2014kxa}.
The observations of the perihelion shifts of not only Mercury but also Venus, Earth, and Mars are now analysed through the dynamical models of planets, large asteroids, and the massive ring of small asteroids in the Solar System in the post-Newtonian approximation of general relativity within considerably high accuracy (e.g., Ref.~\cite{Pitjeva:2005}).
Recently, this classical test has again attracted attention because 
a number of very massive and very compact objects that are 
candidates for black holes
have been observed and accessible by observations of stellar orbits, 
shadows, and gravitational waves.
In particular, the general relativistic periapsis shift of a star called S2, 
whose periapsis distance is $\sim 10^{3}$ times the gravitational radius of
the central supermassive compact object at Sagittarius A* (Sgr A*), 
has been observed by Gravity Collaboration~\cite{Abuter:2020dou}.
Other stars having even closer orbits
to the central object have been reported~\cite{Peissker:2022}.

Although Kerr black holes with almost vacuum surroundings 
remain a standard assumption for the central objects, 
a lot of alternative possibilities have been discussed because such a strong 
field definitely is a new frontier of gravitational physics in this century.
As for the nature of the central objects themselves,
it has been discussed that
they might not be black holes but dense 
cores~\cite{Ruffini:2014zfa, Arguelles:2021jtk, Igata:2022nkt}, 
boson stars~\cite{Grould:2017rzz}, 
naked singularities~\cite{Bambhaniya:2021ybs, Ota:2021mub}, 
wormholes~\cite{Manna:2019tpn} or other exotic compact objects.
See Ref.~\cite{Cardoso:2019rvt} for a recent review of dark compact objects 
and references therein.
On the other hand,
since the black hole no-hair conjecture is broken in certain 
circumstances (e.g., Refs.~\cite{Herdeiro:2015waa,Herdeiro:2015gia}), it is impossible 
without detailed observation to exclude such a possibility as
the central massive compact object may be a black hole 
with a significant hair. 
Constraints on the hairs of the central object M87* have been obtained by 
Event Horizon Telescope (EHT)~\cite{EventHorizonTelescope:2021dqv}.
Even if the central object is a standard black hole well approximated 
by a Kerr spacetime, it might be surrounded by a dark matter 
spike~\cite{Bertone:2009kj} or any other fields.
It is also discussed that gravitational theory is significantly modified from Einstein gravity in such a strong field 
regime, so that the black hole is different from Kerr's one, which may be 
realised in some sort of scalar-tensor theories~\cite{Babichev:2016rlq}, 
Chern-Simons gravity, Einstein-Gauss-Bonnet-dilaton 
gravity~\cite{Pani:2009wy}, and so on.
See Ref.~\cite{Barack:2018yly} for a recent review of black holes in 
modified theories of gravity.
In this paper, we collectively refer to the above physical matter fields 
and the effective matter fields due to the modification of gravity
surrounding the central object as {\it dark component} to distinguish it from the 
so-called dark matter.

It has been found that 
periapsis shifts in these nonstandard scenarios can be very different from 
those for the Kerr black hole.
In particular, the possibility of a backward periapsis shift, i.e., 
a periapsis shift in a direction backward to the orbital rotation, 
has been shown near
a dense core~\cite{Arguelles:2021jtk,Igata:2022nkt},
a boson star~\cite{Grould:2017rzz}, 
a naked singularity~\cite{Bambhaniya:2021ybs,Ota:2021mub},
a wormhole~\cite{Manna:2019tpn}, 
and also in dark matter distribution around a standard 
black hole~\cite{Igata:2022rcm,Jusufi:2022jxu}
in more or less astrophysically reasonable conditions. 
In these studies, the observed value of the periapsis shift 
has to be interpreted model by model. Although 
this is a standard method, what the observational data exactly mean 
on the spacetime geometry may not be so clear.

On the other hand, to interpret the backward periapsis shift, 
what can be called the `extended-mass effect' on the periapsis 
shift should play a key role. This effect has been discovered in Newtonian gravity 
by Jiang and Lin (1984)~\cite{Jiang:1984}, who showed that
the periapsis shift due to this effect is backward.
This effect has also been used to constrain the abundance of dark matter in 
the Solar System~\cite{Gron:1995rn,Khriplovich:2006sq, Khriplovich:2007qt} and 
the Galactic Centre~\cite{Rubilar:2001,Nucita:2007qp,Iwata:2016ivt}.
The amount of the extended mass inside the orbit of S2 in the Galactic Centre 
cannot exceed approximately 0.1\% of 
the central mass in Ref.~\cite{Abuter:2020dou} and
is also estimated to be less 
than 0.5 \% of the mass of the central compact object~\cite{Takamori:2020ntj}.
A detailed analysis and forecasts for future measurements on the extended 
dark matter in the orbit of S2 were implemented in Ref.~\cite{Heissel:2021pcw}.

A very powerful tool free from specific modellings
is the parameterised post-Newtonian approach~\cite{Weinberg:1972}, 
in which 
the metric functions are expanded as power series in 
the small parameter $GM/(c^{2}r)$ with $G$, $c$, $M$, and $r$ being 
the gravitational constant, the speed of light,  
the Arnowitt-Deser-Misner mass, and the radial coordinate, respectively. 
In other papers such as Ref.~\cite{Rubilar:2001}, the (parameterised) post-Newtonian Lagrangian is used to derive the periapsis shift by assuming some trial functions for dark matter distribution.

In the present paper, 
we first derive formulae for the periapsis shift of a 
quasi-circular orbit in a general static spherically symmetric spacetime
without invoking a weak-field approximation, 
so that we can largely extend the above mentioned 
Newtonian extended-mass effect
to the full-order effects of dark component.
In fact, we do not take any specific modellings except for spherical symmetry.
With a general metric in spherically symmetric spacetimes, 
without assuming any model metric or any matter fields or any gravitational field equations, we derive formulae for the periapsis shift in terms of the Einstein tensor and either 
the gravitational mass or the orbital frequency. 
Therefore, the current method enables 
us to directly translate the observational data to the constraint on the spacetime 
curvature
on the orbit of the star, independent from physics which the background spacetime relies on and free from the ansatz about spatial matter distribution. 
Thus, the current approach is novel from the previous approaches for 
the periapsis shift studies, in which some specific spacetimes, such as the Schwarzschild spacetime~\cite{Poisson:2014} and the Janis-Newman-Winicour 
spacetime~\cite{Ota:2021mub}, and/or some approximations 
on the gravitational field, such as the weak-field one~\cite{Pati:2000vt,Pati:2002ux}, 
are adopted.
We next formulate several physically interesting expansions and discuss their 
applications. In particular, we show that in the weak-field regime of general relativity, 
the active gravitational mass density plays a key role.
See Ref.~\cite{Schutz:2022} for the active gravitational mass density.
If the periapsis precession is more advanced than that 
in the vacuum spacetime, the strong energy condition must be violated.
See Ref.~\cite{Wald:1984rg} for the strong energy condition.
A matter field which violates this 
condition is dubbed {\it dark energy} in cosmology. 
In fact, the violation of the strong energy condition implies a 
negative active gravitational mass 
density so that the gravitational force it sources is locally repulsive. 
It accelerates the expansion of the universe on the one hand, while 
gives an excessively forward periapsis shift of the quasi-circular orbit on the other.
It should be noted that the present approach assumes that a small body follows a 
geodesic with respect to the background spacetime metric even if it is self-gravitating.

This paper is organised as follows. 
In Sec.~\ref{sec:circular_orbits}, we briefly review circular orbits 
and the periapsis shift of quasi-circular orbits in a general static 
spherically symmetric spacetime 
in terms of the metric functions. In Sec.~\ref{sec:periapsis_shift}, we derive
two exact formulae in terms of the components of the Einstein tensor 
in the sense that the gravitational field is not assumed to be weak.
In Sec.~\ref{sec:limits}, we discuss several different physically interesting expansions
 of the formulae, such as in the post-Newtonian regime, the diffuse regime, and the weak-field regime. 
In Sec.~\ref{sec:discussion}, we demonstrate the applications of the present results 
for the abundance of dark component in the Solar System through the perihelion shifts of the planets and in the Galactic Centre through the periapsis shift of S2. 
Section~\ref{sec:summary} is devoted to conclusion.
We discuss the periapsis shifts in the anti-de Sitter spacetime in Appendix~\ref{sec:ADS},
in the post-Newtonian and diffuse regime in Appendix~\ref{sec:PNNMO}, 
and with special matter fields, such as a vanishing radial stress 
and a perfect fluid, in Appendix~\ref{sec:special_matter_fields}. 
We use units in which $G=1$ and $c=1$ throughout this paper unless otherwise stated.

\section{Preliminaries
\label{sec:circular_orbits}}

Although most of the discussions in this section are given in standard textbooks 
such as~Refs.~\cite{Weinberg:1972,Poisson:2014,Schutz:2022}, we briefly present 
them for later convenience.
The line element in a static spherically symmetric spacetime can be written in the following form:
\begin{equation}
 ds^{2}=-e^{\nu(r)}dt^{2}+e^{\lambda(r)}dr^{2}+r^{2}(d\theta^{2}+\sin^{2}\theta d\phi^{2}),
\end{equation}
where $\nu$ and $\lambda$ are arbitrary functions of $r$. The gravitational mass 
$m$ is defined as 
\begin{equation}
e^{\lambda(r)}=\left(1-\frac{2m(r)}{r}\right)^{-1},
\label{eq:m}
\end{equation}
which implies $r>2m$.

We model an object orbiting in this spacetime by a test particle, whose trajectory is given by a timelike geodesic. The Lagrangian of the test particle is given by 
\begin{equation}
 {\cal L}=\frac{1}{2}\left[-e^{\nu}\dot{t}^{2}+e^{\lambda}\dot{r}^{2}+r^{2}(\dot{\theta}^{2}+\sin^{2}\theta \dot{\phi}^{2})\right],
\end{equation}
where the dot denotes the differentiation with respect to the affine parameter.
By symmetry, we can assume that the orbit is on the $\theta=\pi/2$ plane.
There are conserved quantities, energy $E$ and angular momentum $L$, 
associated with the time-translational Killing vector $\partial_{t}$ 
and the rotational Killing vector $\partial_{\phi}$, respectively.
They are related to $\dot{t}$ and $\dot{\phi}$ as follows:
\begin{equation}
\dot{t}=e^{-\nu}E, \quad \dot{\phi}=\frac{L}{r^{2}}.
\label{eq:dott_dotphi}
\end{equation}
The Euler-Lagrange equations then reduce to 
\begin{equation}
 \ddot{r}+V'(r)=0,
\label{eq:ddotr}
\end{equation}
where the prime denotes the differentiation with respect to $r$,
and the normalisation condition $g_{\mu\nu}\dot{x}^{\mu}\dot{x}^{\nu}=-1$ 
implies 
\begin{equation}
 \frac{1}{2}\dot{r}^{2}+V(r)=0
\label{eq:dotr^2}
\end{equation}
with $V(r)$ being the effective potential given by 
\begin{eqnarray}
 V(r)=\frac{1}{2}e^{-\lambda}\left[\left(1+\frac{L^{2}}{r^{2}}\right)-e^{-\nu}E^{2}\right]. 
\label{eq:V}
\end{eqnarray}
The affine parameter coincides with the proper time $\tau$.

Let us first concentrate on circular orbits at $r=r_{0}$.
From Eqs.~(\ref{eq:ddotr}) and (\ref{eq:dotr^2}), 
we have $V=V'=0$ at $r=r_{0}$. 
Then, we can deduce~\cite{Cardoso:2008bp}
\begin{equation}
 V''=e^{-\lambda}\left(
3\frac{L^{2}}{r^{4}}+\frac{\nu'' - \nu'^{2}}{2}e^{-\nu}E^{2}\right)
\label{eq:Vc''}
\end{equation}
at $r=r_{0}$, 
where and hereafter we write $r$ for $r_{0}$ for brevity unless it is misleading.
Equation~(\ref{eq:V}) 
with $V(r)=V'(r)=0$ implies
\begin{eqnarray}
 E^{2}= \frac{2e^{\nu}}{2-r\nu'}, \quad 
 L^{2}= \frac{r^{3}\nu'}{2-r\nu'}.
\label{eq:L2nu'}
\end{eqnarray}
Since $E^{2}>0$ and $L^{2}>0$, the condition $0<r\nu'<2$ 
must be satisfied for the existence of the circular orbit.
Substituting the above into Eq.~(\ref{eq:Vc''}), we find
\begin{equation}
 V''=\frac{3r^{-1}\nu' +\nu''-(\nu')^{2}}{2-r\nu'}e^{-\lambda}.
\label{eq:V''c_metric}
\end{equation}
The circular orbit is stable, unstable, and marginally stable if $V''$
is positive, negative, and zero, respectively.

If a stable circular orbit at $r=r_{0}$ with $V''>0$ is perturbed 
as $r=r_{0}+\delta r$ with
$\delta r$ being infinitesimally small, 
it becomes a quasi-circular orbit, where
$\delta r$ obeys a simple harmonic motion
as understood in Eq.~(\ref{eq:ddotr}).
The orbital angular velocity and radial frequency of the particle 
in terms of the proper time are given by
\begin{eqnarray}
\omega_{\phi}=\dot{\phi}
=\frac{L}{r^{2}}=\sqrt{\frac{\nu'}{r(2-r\nu')}} 
\label{eq:omegaphi}
\end{eqnarray}
and 
\begin{eqnarray}
 \omega_{r}=\sqrt{V''},
\label{eq:omegar}
\end{eqnarray}
respectively, where Eq.~(\ref{eq:L2nu'}) has been used, 
and Eq.~(\ref{eq:omegar}) comes from Eq.~(\ref{eq:ddotr}) under the perturbation $r=r_{0}+\delta r$.
Using Eqs.~(\ref{eq:V''c_metric}), (\ref{eq:omegaphi}), and (\ref{eq:omegar}), 
the periapsis shift $\Delta \phi_{p}$ 
can be calculated to give~\cite{Mak:2018hcn} 
\begin{equation}
 \Delta\phi_{p}=2\pi\left(\frac{\omega_{\phi}}{\omega_{r}}-1\right)=2\pi \left(\frac{1}{\sqrt{A}}-1\right),
\label{eq:Deltaphip_A}
\end{equation}
where 
\begin{eqnarray}
 A:= r e^{-\lambda}\frac{\nu''-(\nu')^{2}+\frac{3}{r}\nu'}{\nu'}.
\label{eq:def_A}
\end{eqnarray}
The function $A$ is convenient for further computation.
From Eq.~(\ref{eq:Deltaphip_A}), we can see that 
the shift is forward, backward, and zero,
if $0<A<1$, 
$A>1$, and $A=1$, 
respectively.
Note that $A>0$ is guaranteed by the assumption of the stable orbit $V''>0$.

In the case of the Schwarzschild spacetime, where 
\begin{equation}
 e^{\nu}=1-\frac{2M}{r}, \quad e^{\lambda}=\left(1-\frac{2M}{r}\right)^{-1}, 
\end{equation}
Eqs.~(\ref{eq:L2nu'}) and (\ref{eq:omegaphi}) imply
\begin{equation}
 M=\frac{r^{3}\omega_{\phi}^{2}}{1+3r^{2}\omega_{\phi}^{2}}.
\label{eq:M_omega_phi_Sch}
\end{equation}
Thus, the expression for $A$ in the Schwarzschild spacetime reduces to
\begin{equation}
 A=1-\frac{6M}{r}=\frac{1-3r^{2}\omega_{\phi}^{2}}{1+3r^{2}\omega_{\phi}^{2}},
\label{eq:A_Sch}
\end{equation}
and, hence, 
we reproduce the following expressions~\cite{Poisson:2014,Walters:2018,Tucker:2018rgy}
\begin{equation}
\Delta\phi_{p}=2\pi \left(\frac{1}{\sqrt{1-\frac{6M}{r}}}-1\right)
=2\pi \left(
\sqrt{\frac{1+3r^{2}\omega_{\phi}^{2}}{1-3r^{2}\omega_{\phi}^{2}}}-1
\right).
\label{eq:Deltaphip_Sch}
\end{equation}
Equation~(\ref{eq:Deltaphip_Sch}) 
reproduces the well-known 
expressions 
\begin{eqnarray}
\Delta\phi_{p}\simeq \frac{6\pi M}{r}\simeq 6\pi r^{2}\omega_{\phi}^{2}
\label{eq:well-known_formula}
\end{eqnarray}
in the weak-field limit.

\section{Exact formulae for the periapsis shift \label{sec:periapsis_shift}}

\subsection{Expression in terms of the Einstein tensor and the gravitational mass}
Let $\{\vec{e}_{\hat{\alpha}}\}_{\hat{\alpha}=0,1,2,3}$ be a natural tetrad 
basis given by 
\begin{equation}
 \vec{e}_{\hat{0}}=e^{-\nu/2}\frac{\partial}{\partial t},\quad
 \vec{e}_{\hat{1}}=e^{-\lambda/2}\frac{\partial}{\partial r},\quad 
 \vec{e}_{\hat{2}}=\frac{1}{r}\frac{\partial}{\partial \theta},\quad 
 \vec{e}_{\hat{3}}=\frac{1}{r\sin\theta}\frac{\partial}{\partial \phi}.
\end{equation}
The tetrad components of the Einstein tensor, $G_{\hat{\alpha}\hat{\beta}}=e^{~\mu}_{\hat{\alpha}}e^{~\nu}_{\hat{\beta}}G_{\mu\nu}$,  
can be calculated to give
\begin{eqnarray}
G_{\hat{0}\hat{0}}&=& \frac{1}{r^{2}}[r(1-e^{-\lambda})]', 
\label{eq:G00}\\
G_{\hat{1}\hat{1}}&=&\frac{e^{-\lambda}}{r}\nu'-\frac{1}{r^{2}}(1-e^{-\lambda}), 
\label{eq:G11}\\
G_{\hat{2}\hat{2}} &=& G_{\hat{3}\hat{3}}=\frac{1}{2}e^{-\lambda}\left(\nu''+\frac{\nu'^{2}}{2}-\frac{\nu'\lambda'}{2}+\frac{\nu'-\lambda'}{r}\right), 
\label{eq:G22}
\end{eqnarray} 
while all other components of $G_{\hat{\alpha}\hat{\beta}}$ vanish.
The contracted Bianchi identity 
$\nabla_{\mu}G^{\mu\nu}=0$ 
implies
\begin{equation}
 G_{\hat{1}\hat{1}}'+\frac{1}{2}\nu' (G_{\hat{0}\hat{0}}+G_{\hat{1}\hat{1}})
  +\frac{2}{r}(G_{\hat{1}\hat{1}}-G_{\hat{2}\hat{2}})=0.
\label{eq:contracted_Bianchi}
\end{equation}

Equation~\eqref{eq:G22}
implies
\begin{equation}
 \nu''=2e^{\lambda}G_{\hat{2}\hat{2}}-\frac{(\nu')^{2}}{2}+\frac{\nu'\lambda'}{2}-\frac{\nu'-\lambda'}{r},
\label{eq:nu''}
\end{equation}
while Eqs.~(\ref{eq:G00}) and (\ref{eq:G11}) imply
\begin{eqnarray}
 \lambda'&=& r e^{\lambda}\left[G_{\hat{0}\hat{0}}-\frac{1}{r^{2}}(1-e^{-\lambda})\right], 
\label{eq:lambda'} \\
 \nu'&=& re^{\lambda}\left[G_{\hat{1}\hat{1}}+\frac{1}{r^{2}}(1-e^{-\lambda})\right],
\label{eq:nu'}
\end{eqnarray}
respectively.

Since Eqs.~(\ref{eq:m}) and (\ref{eq:nu'}) imply
\begin{equation}
 \nu'=r\frac{\frac{2m}{r^{3}}+G_{\hat{1}\hat{1}}}{1-\frac{2m}{r}}, 
\label{eq:nu'_m}
\end{equation}
the conditions $E^{2}>0$ and $L^{2}>0$ or $0<r\nu'<2$ are met if and only if 
\begin{equation}
-\frac{2m}{r^{3}}<G_{\hat{1}\hat{1}}<\frac{2}{r^{2}}\left(1-\frac{3m}{r}\right)
\label{eq:E2L2condition}
\end{equation}
is satisfied.

From Eqs.~(\ref{eq:def_A}), (\ref{eq:nu''}), (\ref{eq:lambda'}), and (\ref{eq:nu'}), we obtain
$A=A_{m 0}+A_{m 1}$, where
\begin{eqnarray}
 A_{m 0}&=& 1-\frac{6m}{r}, \label{eq:Am0}\\
 A_{m 1}&=& \left(1-\frac{2m}{r}\right)\frac{G_{\hat{0}\hat{0}}+G_{\hat{1}\hat{1}}+2G_{\hat{2}\hat{2}}}{\frac{2m}{r^{3}}+G_{\hat{1}\hat{1}}}+\frac{1}{2}
(G_{\hat{0}\hat{0}}-3G_{\hat{1}\hat{1}})r^{2}.
\label{eq:Am1}
\end{eqnarray}
It is clear that for the locally Ricci-flat (or `vacuum' for brevity) spacetime,
this expression reproduces the Schwarzschild formula (\ref{eq:A_Sch})
or (\ref{eq:Deltaphip_Sch}) 
with the identification $M=m(r)$.
The shift is more and less forward 
than that in the vacuum, 
if and only if 
$A_{m 1}$ is negative and positive, respectively. 
A spacetime with $A_{m1}=0$ gives the periapsis shift identical 
with that in the vacuum spacetime.
What we have shown is that under 
the assumption of the geodesic motion, the 
periapsis shift is determined by the gravitational mass and the Einstein tensor at 
the radius of the orbit.

To demonstrate the vast applicability of the exact formula obtained above, 
we apply it to maximally symmetric spacetimes.
This is delegated to Appendix~\ref{sec:ADS}.

\subsection{Expression in terms of the Einstein tensor and the orbital angular velocity \label{sec:omega_phi}}

In Newtonian gravity, the angular velocity of the circularly orbiting star is used 
to estimate the mass of the central gravitating object through Kepler's third law.
This is also possible in the Schwarzschild spacetime as is seen 
in Eq.~(\ref{eq:M_omega_phi_Sch}). However, in the general case, 
from Eqs.~(\ref{eq:L2nu'}), (\ref{eq:omegaphi}), and (\ref{eq:nu'_m}), 
we obtain 
\begin{equation}
 m=\frac{r^{3}\omega_{\phi}^{2}-\frac{1}{2}G_{\hat{1}\hat{1}} r^{3}(1+r^{2}\omega_{\phi}^{2})}
{1+3r^{2}\omega_{\phi}^{2}}.
\end{equation}
Therefore, without knowing $G_{\hat{1}\hat{1}}$, the orbital 
angular velocity $\omega_{\phi}$
is not enough to determine $m(r)$ even if we know $r$. In other words, $\omega_{\phi}$ is more accessible than $m$.
Thus, it would be more useful to have the expression for $A$ in terms of $\omega_{\phi}$ and $r$.

The result is written in the form $A=A_{\omega 0}+A_{\omega 1}$, where 
\begin{eqnarray}
 A_{\omega 0}&=&
\frac{1-3r^{2}\omega_{\phi}^{2}}{1+3r^{2}\omega_{\phi}^{2}},
\label{eq:Aomega0}\\
A_{\omega 1}&=&
\frac{1}{\omega_{\phi}^{2}}
\left[
\frac{1}{2}
\left(G_{\hat{0}\hat{0}}+2G_{\hat{2}\hat{2}}+\frac{1+7 r^{2}\omega_{\phi}^{2}}
{1+3r^{2}\omega_{\phi}^{2}}G_{\hat{1}\hat{1}}\right)
+r^{2}\omega_{\phi}^{2}(G_{\hat{0}\hat{0}}+G_{\hat{2}\hat{2}})
\right].
\label{eq:Aomega1}
\end{eqnarray}
The first term 
$A_{\omega 0}$ has the form identical to $A$ 
for the Schwarzschild spacetime, while the second term  
$A_{\omega 1}$ denotes the deviation from it. 
If $A_{\omega 1}=0$, the periapsis shift is identical to that 
for the Schwarzschild spacetime with the same $\omega_{\phi}$ and $r$.
The shift is more and less forward if $A_{\omega 1}$ is negative and 
positive, respectively.
It would be interesting that the exact expression (\ref{eq:Aomega1}) for $A_{\omega 1}$ is linear
 with respect to the Einstein tensor, whereas this is not the case for $A_{m 1}$ 
as is seen in Eq.~(\ref{eq:Am1}).

It should be noted that the angular velocity with respect to the proper time, $\omega_{\phi}=d\phi/d\tau$, is different from the angular velocity measured at infinity, $\Omega_{\phi}:=d\phi/dt$,
where $e^{\nu(r)}\to 1$ in the limit $r\to \infty$ is assumed.
The latter can be written for the circular orbit as
\begin{equation}
 \Omega_{\phi}=\frac{e^{\nu}}{r^{2}}\frac{L}{E}=e^{\nu/2}\sqrt{\frac{\nu' }{2r}},
\end{equation}
where we have used Eqs.~(\ref{eq:dott_dotphi}) and (\ref{eq:L2nu'}). The relation between 
$\omega_{\phi}$ and $\Omega_{\phi}$ is written as
\begin{equation}
 \omega_{\phi}=(1+z)\Omega_{\phi},
\end{equation}
where $1+z:=dt/d\tau$ is the averaged redshift factor, which is given by 
\begin{equation}
 1+z=e^{-\nu}E=e^{-\nu/2}\sqrt{\frac{2}{2-r\nu'}}
\end{equation}
for the circular orbit from Eqs.~(\ref{eq:dott_dotphi}) and (\ref{eq:L2nu'}).
See also Ref.~\cite{Igata:2022rcm} for the discussion on the averaged redshift factor.
Thus, $\omega_{\phi}$ can be calculated from the two observables in principle, 
$\Omega_{\phi}$ and $1+z$.   

\section{Physically interesting limits \label{sec:limits}}

In this section, for simplicity we use the following notation: 
\begin{equation}
 G_{\hat{0}\hat{0}}=8\pi \epsilon,~~ G_{\hat{1}\hat{1}}=8\pi p_{r}, ~~G_{\hat{2}\hat{2}}=G_{\hat{3}\hat{3}}=8\pi p_{t},~~
 m=\frac{4\pi}{3}r^{3}\bar{\epsilon}.
\end{equation}
We also follow the terminology in general relativity, so that 
$\epsilon$, $p_{r}$, $p_{t}$, and $\bar{\epsilon}$
are termed the energy density, radial stress, tangential stress, 
and averaged energy density, respectively.
This is purely for convenience and it is trivial how to translate the current result to 
modified theories of gravity because we never solve the Einstein equation in any case.

\subsection{Post-Newtonian regime \label{sec:PN}}

In the post-Newtonian expansion, assuming $m/r=O(v^{2})$, $r^{2}\omega_{\phi}^{2}=O(v^{2})$, 
$\epsilon/\bar{\epsilon}=O(v^{0})$, $p_{r}/\epsilon=O(v^{2})$, 
and $p_{t}/\epsilon=O(v^{2})$ with $v$ being the speed of the orbiting star,
we obtain
\begin{eqnarray}
\Delta \phi_{p}&=&
-2\pi \left(1-\frac{1}{\sqrt{1+\frac{3\epsilon}{\bar{\epsilon}}}}\right)
+3\pi \left(1+\frac{3\epsilon}{\bar{\epsilon}}\right)^{-3/2}
\left[\frac{m}{r}\left(2+\frac{\epsilon}{\bar{\epsilon}}\right)
-\frac{p_{r}+2p_{t}}{\bar{\epsilon}}+\frac{3\epsilon p_{r}}{\bar{\epsilon}^2}
\right] \nonumber \\
&&+O\left(v^{4}\right)
\label{eq:Deltaphip_PN_m}
\\
&=& -2\pi \left[1-\frac{1}{\sqrt{1+\frac{4\pi \epsilon}{\omega_{\phi}^{2}}}}\right]
+2\pi \left(1+\frac{4\pi\epsilon}{\omega_{\phi}^{2}}\right)^{-3/2} 
\left[\left(3-\frac{4\pi \epsilon}{\omega_{\phi}^{2}}\right)r^{2}\omega_{\phi}^{2}
-2\pi \frac{p_{r} +2p_{t}}{\omega_{\phi}^{2}}  \right]\nonumber \\
&&+O(v^{4}),
\label{eq:Deltaphip_PN_omega}
\end{eqnarray}
where we have used Eqs.~(\ref{eq:Am0}), (\ref{eq:Am1}), (\ref{eq:Aomega0}), and (\ref{eq:Aomega1}), and it turns out that $m>0$ must be satisfied.
In the bottom expression, for example, $\epsilon/\omega_{\phi}^{2}$ is understood as
the nondimensional quantity $G\epsilon/(\omega_{\phi}c)^{2}$ 
if $c$ and $G$ are recovered.  
In both of the above expressions, 
the first and second terms on the right-hand side stand for the 
Newtonian and the first post-Newtonian ones, respectively.
The first term is negative if $\epsilon>0$ 
and therefore gives a negative contribution to $\Delta \phi_{p}$. 
This corresponds to the Newtonian extended-mass effect.
As we can see, $\epsilon$, $p_{r}$, and $p_{t}$ also 
affect the second term of the post-Newtonian order.
On the other hand, if $\epsilon=p_{r}=p_{t}=0$ and $m=M$, 
the Newtonian terms vanish, and the post-Newtonian terms reproduce 
the well-known weak-field formulae~(\ref{eq:well-known_formula}) 
for the Schwarzschild spacetime.

\subsection{Diffuse regime \label{sec:NMO}}
Next, we instead assume that 
$|\epsilon|, |p_{r}|, |p_{t}| \ll |\bar{\epsilon}|\sim \omega_{\phi}^{2}$. 
This limit applies if dark component is diffuse near a 
massive central gravitating object.
The periapsis shift is then calculated to give
\begin{eqnarray}
 \Delta \phi_{p}&=&2\pi 
\left(\frac{1}{\sqrt{1-\frac{6m}{r}}}-1\right) \nonumber \\
&&-3\pi \left(1-\frac{6m}{r}\right)^{-3/2}\left(\frac{\epsilon+p_{r}+2p_{t}}{\bar{\epsilon}}-\frac{m}{r}\frac{\epsilon+5p_{r}+4p_{t}}{\bar{\epsilon}}\right)+O(\alpha^{2}) 
\label{eq:Deltaphip_RF_m}\\
&=& 
2\pi \left(\sqrt{\frac{1+3r^{2}\omega_{\phi}^{2}}{1-3r^{2}\omega_{\phi}^{2}}}-1\right)
\nonumber \\
&&-4\pi^{2} \left(\frac{1+3r^{2}\omega_{\phi}^{2}}{1-3r^{2}\omega_{\phi}^{2}}\right)^{3/2}
\left(
\frac{\epsilon +2p_{t}}{\omega_{\phi}^{2}}+\frac{1+7 r^{2}\omega_{\phi}^{2}}
{1+3r^{2}\omega_{\phi}^{2}}\frac{p_{r}}{\omega_{\phi}^{2}}
+2r^{2}\omega_{\phi}^{2}\frac{\epsilon+p_{t}}{\omega_{\phi}^{2}}\right)+O(\alpha^{2}),
\label{eq:Deltaphip_RF_omega}
\end{eqnarray}
where we assume that all of $\epsilon/\bar{\epsilon}$, $p_{r}/\bar{\epsilon}$, and 
$p_{t}/\bar{\epsilon}$ are of the order of $\alpha (\ll 1)$, and it turns out that $m>0$ must hold.
In the above, the first terms on the right-hand side are 
of $O(\alpha^{0})$ and coincide with Eq.~(\ref{eq:Deltaphip_Sch}), 
the Schwarzschild formulae,
while the second terms are of $O(\alpha)$.

\subsection{Weak-field and diffuse regime \label{sec:WFNMO}}
Here we assume that the gravitational field is weak as well as 
dark component is diffuse,
i.e., $|2m/r|=O(h) $ with $h\ll 1$ in addition to $\alpha\ll 1$ in Sec.~\ref{sec:NMO}. 
However, we do not regard
$p_{r}$ and $p_{t}$ much smaller than $\epsilon$, 
so we assume $p_{r}/\epsilon=O(h^{0})$ and $p_{t}/\epsilon=O(h^{0})$
as the nature of dark component is not well-known.
Then, we can expand $\Delta \phi_{p}$ in powers of both $\alpha$ and $h$ as follows:
\begin{eqnarray}
 \Delta \phi_{p}&=&3\pi \left(\frac{2m}{r}-\frac{\epsilon+p_{r}+2p_{t}}{\bar{\epsilon}}\right)
+\frac{27\pi m^{2}}{r^{2}}
-24\pi \frac{m}{r}\frac{\epsilon+\frac{1}{2}p_{r}+\frac{7}{4}p_{t}}{\bar{\epsilon}}
\nonumber \\
&& +O\left(h^{3}\alpha^{0}, h^{2}\alpha^{1} , h^{0}\alpha^{2}\right)
\label{eq:Deltaphip_WFRF_1}
\\
&=&3\pi \left(2r^{2}\omega_{\phi}^{2}-\frac{4\pi }{3}\frac{\epsilon+p_{r}+2p_{t}}{\omega_{\phi}^{2}}\right) 
+9\pi r^{4}\omega_{\phi}^{4}
-44\pi^{2}r^{2}\omega_{\phi}^{2}\frac{\epsilon+\frac{13}{11}p_{r}+\frac{20}{11}p_{t}}
{\omega_{\phi}^{2}} \nonumber \\
&&+O\left(h^{3}\alpha^{0}, h^{2}\alpha^{1} , h^{0}\alpha ^{2}\right),
\label{eq:Deltaphip_WFRF_2}
\end{eqnarray}
where the first terms include those of $O(h^{1}\alpha^{0})$ and $O(h^{0} \alpha^{1})$, 
while the second and third terms are of higher order.
In this regime, the periapsis shift in the lowest orders is forward (backward) if 
${2m}/{r}>(<){\rho_{A}}/{\bar{\rho}}$, i.e.,   
\begin{equation}
 \rho_{A}<(>)\rho_{c}:=\frac{2Gm}{c^{2}r}\bar{\rho}=\frac{3Gm^{2}}{2\pi c^{2}r^{4}}\simeq \frac{3r^{2}\omega_{\phi}^{4}}{2\pi Gc^{2}},
\label{eq:epsilon_c}
\end{equation}
where $c$ and $G$ are recovered, $\bar{\rho}:=\bar{\epsilon}/c^{2}$ is the averaged mass density,  and 
\begin{equation}
\rho_{A}:=\frac{\epsilon+p_{r}+2p_{t}}{c^{2}} 
\label{eq:rho_A}
\end{equation}
is the active gravitational mass density~\cite{Schutz:2022}.
The critical mass density $\rho_{c}$ defined above is estimated to be
\begin{equation}
\rho_{c}\simeq 2.8\times 10^{-15} \mbox{g}/\mbox{cm}^{3} \left(\frac{m}{M_{\odot}}\right)^{2}\left(\frac{r}{\mbox{au}}\right)^{-4}.
\end{equation}
The shift vanishes in the lowest orders if $\rho_{A}=\rho_{c}$.
As we can see in Eqs.~(\ref{eq:Deltaphip_WFRF_1})
and (\ref{eq:Deltaphip_WFRF_2}), if $\Delta \phi_{p}>6\pi m/r\simeq 6\pi r^{2}\omega_{\phi}^{2}$, then, 
$\rho_{A}<0$ must hold, resulting in the violation of the strong energy condition in the sense of general relativity. 
We can rewrite Eqs.~(\ref{eq:Deltaphip_WFRF_1}) 
and (\ref{eq:Deltaphip_WFRF_2}) 
in the lowest orders as
\begin{eqnarray}
 \Delta \phi_{p}&=& \frac{6\pi m}{r}
\left(1-\frac{\rho_{A}}{\rho_{c}}\right)+O\left(h^{2}\alpha^{0}, h^{1}\alpha^{1}, h^{0}\alpha^{2}\right) \nonumber \\
&=&6\pi r^{2}\omega_{\phi}^{2}\left(1-\frac{\rho_{A}}{\rho_{c}}\right)+O\left(h^{2}\alpha^{0}, h^{1}\alpha^{1}, h^{0}\alpha^{2}\right).
\label{eq:Deltaphip_WFRF_epsilon_c}
\end{eqnarray}
In Newtonian gravity, $\rho_{A}$ is replaced by the mass density $\rho$. 
Therefore, we can interpret the result as 
the extended-matter effect in general relativity comes from the gravitational force 
sourced by the active gravitational mass density.

Note that this regime does not mean the Newtonian one.
That is, how diffuse the dark component is, the energy 
density does not necessarily dominate the stress-energy tensor. 
The obvious example is the Reissner-Nordstr\"om black hole.
In this case, the stress-energy tensor of the U(1) gauge field is traceless 
with $\epsilon:p_{r}:p_{t}=1:-1:1$~\cite{Poisson:2009pwt} and thus $\rho_{A}=2\epsilon/c^{2}$. 
The EHT results for M87* have been used to constrain 
the charge of the hypothetical Reissner-Nordstr\"om black 
hole~\cite{EventHorizonTelescope:2021dqv}.
We would also discuss scalar fields because black holes with a scalar hair
have been extensively studied~(e.g., Refs.~\cite{Herdeiro:2015waa,Herdeiro:2015gia}).
A massless scalar field has $\epsilon:p_{r}:p_{t}=1:1:-1$
and thus $\rho_{A}=0$, which gives no anomaly in the lowest order.
The EHT results 
have been also used for the Janis-Newman-Winicour naked singularity~\cite{Janis:1968zz}, an exact solution of the Einstein equation with a massless scalar field, 
to constrain the scalar hair of M87*~\cite{EventHorizonTelescope:2021dqv}, 
while the observational data of S2's orbit
have been used to test the same solution as a model 
for Sgr A*~\cite{Bambhaniya:2022xbz}.
If the stress-energy tensor of a scalar field is dominated by its potential term, 
it has $\epsilon:p_{r}:p_{t}=1:-1:-1$ and therefore $\rho_{A}=- 2 \epsilon/c^2$,
which gives a positive anomaly beyond the Schwarzschild value.

On the other hand, if we additionally assume $|p_{r}|, |p_{t}| \ll |\epsilon|$, we obtain the post-Newtonian and diffuse approximation, 
which implies $\rho_{A}\simeq \epsilon/c^{2}$ in the lowest order.
For completeness, we present the corresponding expression
in Appendix~\ref{sec:PNNMO}.
We also delegate the expressions for the cases of 
a vanishing radial stress and a perfect fluid to 
Appendix~\ref{sec:special_matter_fields}.

\section{Discussion \label{sec:discussion}}

Here, we discuss the implications of the results obtained in the previous section 
for the Solar System and the Galactic Centre. It should be, however, noted that 
the purpose of this section is not to make the constraints 
more accurate than before
but to demonstrate with a simpler analysis 
that we can discuss the range of the active gravitational mass density
from the periapsis shift observation. 

\subsection{Dark component in the  Solar System}

Let us introduce
 \begin{equation}
 \Delta \phi_{p,\mathrm{vac}}:=2\pi \left(\frac{1}{\sqrt{A_{\omega 0}}}-1\right)
= 2\pi \left(\sqrt{\frac{1+3r^{2}\omega_{\phi}^{2}}{1-3r^{2}\omega_{\phi}^{2}}}-1\right)
 \end{equation}
as the periapsis shift in the 
spacetime with vanishing Einstein tensor on the orbit.
In the Solar System, we can reasonably assume that the weak-field and diffuse 
regime is applicable. In this case, as discussed in Sec.~\ref{sec:WFNMO}, 
the periapsis shift for an elliptical orbit, $\Delta \phi_{p, \mathrm{vac}}$, 
and its rate per unit time, $\omega_{p, \mathrm{vac}}$,
are given by
\begin{equation}
  \Delta\phi_{p,\mathrm{vac}}\simeq 6\pi \frac{\omega_{\phi}^{2}a^{2}}{c^{2}(1-e^{2})}
\end{equation}
and 
\begin{equation}
  \omega_{p,\mathrm{vac}}\simeq \frac{3\omega_{\phi}^{3}a^{2}}{c^{2}(1-e^{2})},
\label{eq:perihelion_shift_rate}
\end{equation}
respectively, where $a$ and $e$ are the semimajor axis and eccentricity, 
respectively, $c$ and $G$ are recovered, 
the correction due to eccentricity is included according to 
Refs.~\cite{Einstein:1916vd,Weinberg:1972},
and $\omega_{\phi}=2\pi/T_{\phi}$ with $T_{\phi}$ being the orbital period.
In the weak-field and diffuse regime, 
the effect of nonvanishing eccentricity can be included by the method of 
osculating orbital elements. See Ref.~\cite{Poisson:2014} and references therein.
With an additional assumption of nearly homogeneous matter 
distribution over the orbit, 
a totally parallel calculation with that in Ref.~\cite{Iwata:2016ivt} 
recasts Eq.~(\ref{eq:Deltaphip_WFRF_epsilon_c}) 
to the following form~\cite{Yoo:2023}:
\begin{eqnarray}
\Delta\phi_{p}&\simeq& 6\pi \frac{\omega_{\phi}^{2}a^{2}}{c^{2}(1-e^{2})}
\left[1-(1-e^{2})^{3/2}\frac{\rho_{A}}{\rho_{c}}\right], 
\label{eq:deltaphip_A}
\\
\omega_{p}&\simeq& \frac{3\omega_{\phi}^{3}a^{2}}{c^{2}(1-e^{2})}
\left[1-(1-e^{2})^{3/2}\frac{\rho_{A}}{\rho_{c}}\right],
\label{eq:omegap_A}
\end{eqnarray}
where $\rho_{A}$ is the active gravitational mass density defined in 
Eq. (\ref{eq:rho_A}) and $\rho_{c}$ is the critical density defined in 
Eq.~(\ref{eq:epsilon_c}) but in term of the semimajor axis $a$ in place of $r$.

If we take the semimajor axis $a$ and the eccentricity $e$ 
for the orbit of Mercury as $a\simeq 0.387$ au and $e\simeq 0.206$,
we obtain 
\begin{eqnarray}
 \rho_{c}&\simeq& 1.2\times 10^{-13} \left(\frac{m}{M_{\odot}}\right)^{2}\left(\frac{a}{0.387 \mbox{au}}\right)^{-4} \mbox{g}/\mbox{cm}^{3}, \\
 \rho_{A} &\simeq & (1.1 \pm 1.5) \times 10^{-17} \left(\frac{m}{M_{\odot}}\right)^{2}\left(\frac{a}{0.387 \mbox{au}}\right)^{-4} \mbox{g}/\mbox{cm}^{3},
\end{eqnarray}
where we have used the difference $\Delta \omega_{p}=-0.0036\pm 0.0050 ''/\mbox{cy}$ from Table 3 in Ref.~\cite{Pitjeva:2005} with $\Delta \omega_{p}:=\omega_{p}-\omega_{p,\mathrm{vac}}$. 
A more conservative estimate $\omega_{p}=42.98\pm 0.24 ''/\mbox{cy}$ 
in Ref.~\cite{Nobili:1986} yields 
$\rho_{A}=(0.0 \pm 7.4) \times 10^{-16} $ g$/$cm$^{3}$. 
The active gravitational mass densities $\rho_{A}$ of dark component for the orbits of Mercury, Venus, Earth, and Mars calculated from data in Ref.~\cite{Nobili:1986,Pitjeva:2005} are summarised in Table~\ref{table:planets}.
See also Refs.~\cite{Biswas:2008cw,Wilhelm:2014,Will:2014kxa} 
for possible discrepancies in the perihelion shifts for the solar planets.
We can see from Table~\ref{table:planets} that 
the active gravitational mass density of dark component 
is much more severely constrained 
on Earth's and Mars' orbits than on Mercury's and Venus' ones.
The value only for Venus is in the range of the violation of the strong energy condition, 
although the planetary ephemerides model is less accurate for Venus 
than other planets~\cite{Pitjeva:2005,Khriplovich:2006sq}. 
Note that the values are much higher than that of Galactic dark matter $\sim 10^{-24} \mbox{g}/\mbox{cm}^{3}$ needed to explain 
the flat rotation curve of the Galaxy~\cite{Nesti:2013uwa}, although the latter 
does not exclude the locally much higher density of dark component in the Solar System.
It is suggested that the theoretical uncertainty may overcome the contribution of the background dark matter~\cite{Fienga:2023ocw}.
\begin{table}
\caption{Active gravitational mass density of dark component on the orbits of the planets: $a$ and $e$ are the semimajor axis and the eccentricity, respectively, and $\omega_{p,\mathrm{vac}}$ and 
$\Delta \omega_{p}$ denote the perihelion shift 
calculated by Eq.~(\ref{eq:perihelion_shift_rate})
and the difference between the actual observation and the model one, respectively. 
The rightmost column shows 
the observational constraints on the
active gravitational mass density 
$\rho_{A} $ 
obtained in the current paper.
The values for $a$ and $e$ are taken from Ref.~\cite{JPLAPP}.
The values for the orbital period $T_{\phi}$ are taken from Ref.~\cite{JPLPPP}.
The values for $\Delta \omega_{p}$ are taken from the bottom row of Table 3 in Ref.~\cite{Pitjeva:2005} except for that in the second row that is taken from Ref.~\cite{Nobili:1986}. The values for $\Delta \omega_{p}$ are obtained only after extensive elaborated calculations.
\label{table:planets}}
\begin{tabular}{|c||c|c|c|c|c|c|}
\hline 
Planet & $a$ [au]& $e$ & $T_{\phi}$ [y] & $\omega_{p, \mathrm{vac}}$ [$''/$cy] & $\Delta \omega_{p}$ [$''/$cy] & $\rho_{A}$ [g$/$cm$^{3}$] \\
\hline \hline 
Mercury & 0.38709927 & 0.20563593 & 0.2408467 & 42.980781 & $-0.0036\pm 0.0050$ & $(1.1\pm 1.5) \times 10^{-17}$\\
 & & & & & $0.00\pm 0.24$ & $(0.0 \pm 7.4) \times 10^{-16} $ \\
Venus & 0.72333566 & 0.00677672 & 0.61519726 & 8.6247000 & $0.53\pm 0.30 $ & $(-6.3\pm 3.6) \times 10^{-16}$ \\
Earth & 1.00000261 & 0.01671123 & 1.0000174 & 3.8387371 & $-0.0002\pm 0.0004 $ & $(1 \pm 3) \times 10^{-19}$ \\
Mars & 1.52371034 &  0.09339410 &  1.8808476 & 1.3509405 & $0.0001\pm 0.0005$ & $(-0.4\pm 2)\times 10^{-19}$\\
\hline
\end{tabular} 
\end{table}

Note that an upper limit $1.8\times 10^{-16}$ g $/$cm$^{3}$ 
on dark matter mass density is obtained through 
an apparently similar argument for 1566 Icarus orbit by Gr\o n and Soleng (1996)~\cite{Gron:1995rn}. To do this, they 
obtained a certain form of the metric assuming nonrelativistic dust with a constant density. 
They reported a
positive, probably due to typo though, anomaly in the periapsis shift of the quasi-circular orbit in this spacetime metric and obtained the upper limit of dark matter density using the accuracy $8\%$ in the perihelion shift. 
In reality, the density of dark component should 
naturally depend on the distance from the Sun.
The current approach can consistently incorporate such spatial dependence
without assuming the nature of dark component, while the model in Ref.~\cite{Gron:1995rn} cannot.

We should also note that the constraints on the active gravitational mass density of 
dark component obtained in the current paper are comparable with those on the mass density obtained in Khriplovich and 
Pitjeva (2006)~\cite{Khriplovich:2006sq} and 
Khriplovich (2007)~\cite{Khriplovich:2007qt}. 
The difference is that the present approach is free from assumptions about 
the nature of dark component except for spherical symmetry, 
while in Refs.~\cite{Khriplovich:2006sq} and \cite{Khriplovich:2007qt},
dark matter is assumed to be nonrelativistic dust with 
the constant density and the power-law one, respectively.
In fact, as we have seen, what we can constrain is not the mass density itself 
but {\it the active gravitational mass density} $\rho_{A}$.
Note also that we obtain both the upper and lower limits and thus can discuss the
violation of the strong energy condition, while the above previous papers 
discuss the upper limit only. 

\subsection{Dark component in the Galactic Centre}

Next let us discuss the application of 
the present approach to
S2 star. 
Since the orbit of S2 has 
the semi-major axis $a\simeq 970$ au
and the eccentricity $e\simeq 0.88466 $, 
we can estimate the critical density $\rho_{c}$
there as follows:
\begin{eqnarray}
 \rho_{c} \simeq 9.8\times 10^{-8} M_{\odot}/\mbox{au}^{3}
\left(\frac{m}{4.3\times 10^{6} M_{\odot}}\right)^{2}
\left(\frac{a}{970 \mbox{au}}\right)^{-4}.
\end{eqnarray}
We assume that the weak-field and diffuse regime is valid 
for the S2 orbit. 
Putting
\begin{equation}
 \Delta \phi_{p}=f \Delta \phi_{p,\mathrm{vac}},
\label{eq:def_f}
\end{equation}
we can recast Eq.~(\ref{eq:deltaphip_A}) to
\begin{equation}
\rho_{A}\simeq (1-f)(1-e^{2})^{-3/2}\rho_{c}.
\end{equation}
If we adopt the observational result $f=1.10\pm 0.19$~\cite{Abuter:2020dou}, 
we immediately reach the conclusion
\begin{eqnarray}
 \rho_{A}=(-1.0\pm 1.8) \times 10^{-7} M_{\odot}/\mbox{au}^{3}
\left(\frac{m}{4.3\times 10^{6} M_{\odot}}\right)^{2}
\left(\frac{a}{970 \mbox{au}}\right)^{-4}. 
\label{eq:S2_epsilon_range}
\end{eqnarray}
This value corresponds to that for the active gravitational energy density
$\rho_{A}c^{2}\simeq (-3.2\pm 6.2) \times 10^{10} \mbox{GeV}/\mbox{cm}^{3}$.

This is a rather stringent constraint on the 
matter density in the vicinity of Sgr A*.
In fact, we find that 
the mass within $a\simeq 970~\mbox{au}$ is bounded above by $\simeq 3.7\times 10^{2}M_{\odot}$ by simply multiplying $\rho_{A}$ with the ball's volume. Remarkably, this simple and rough estimate is comparable with 
more robust upper limit estimates by other authors~\cite{Abuter:2020dou,Heissel:2021pcw,Takamori:2020ntj}, where 
the dark matter assumption is adopted with spatial inhomogeneous distribution.
Although strictly speaking, the current method is not completely
justified to highly eccentric orbits as we can not rely on  
the assumption of approximately homogeneous
distribution, we can assume that it gives the same result in order of magnitude 
for moderately eccentric orbits.

Moreover, 
the effect of the spin of the central massive object must be estimated.  
The spinning object gives the deviation of the periapsis shift from that in 
the Schwarzschild 
in the 1.5th post-Newtonian order in terms of the orbital velocity of the stars.
If we simply add this effect to Eq.~(\ref{eq:deltaphip_A}) in the weak-field regime,
we can write it to the following form:
\begin{eqnarray}
\Delta\phi_{p}&\simeq& 6\pi \frac{\omega_{\phi}^{2}a^{2}}{c^{2}(1-e^{2})}
\left[1-(1-e^{2})^{3/2}\frac{\rho_{A}}{\rho_{c}}\mp \frac{4a_{*}}{3(1-e^{2})^{1/2}}
\left(\frac{m}{a}\right)^{1/2}\right], 
\label{eq:deltaphip_A_spin}
\end{eqnarray}
where the orbit is assumed to be on the equatorial plane, 
$a_{*}$ is the nondimensional Kerr parameter of the central body
and the upper and lower signs correspond to the orbits prograde and retrograde to the 
spin, respectively~\cite{Boyer:1965,Bini:2005dy,Vogt:2008zs}.
The relative deviation due to the spin effect is thus $1.9\%$ or less for S2 if $|a_{*}|\le 1$.
This implies that the spin effect alters Eq.~(\ref{eq:S2_epsilon_range})
to $\rho_{A}=(-1.0\pm 2.0)\times 10^{-7}M_{\odot}/\mbox{au}^{3}$, 
while we can in principle distinguish 
the spin effect by observing both prograde and retrograde orbits.
Note that we cannot distinguish between the physical matter fields and the effective ones induced by modified theories of gravity in the present purely kinematical analysis.

In spite of all the above large uncertainties, it
is intriguing that 
although the constraint~(\ref{eq:S2_epsilon_range})
is consistent with $\rho_{A}=0$, 
the best-fit value $f=1.10$ is in the range corresponding to  
the violation of the strong energy condition. 
If the strong energy condition is violated, 
one possibility is that there is a matter field with a positive energy density but 
a large negative stress, i.e., {\it dark energy}. 
Another possibility is that $\epsilon$ itself is negative, implying the violation of 
the weak energy condition.
Although the violation of the weak energy condition is regarded as very serious,  
we should recall that what is directly constrained is 
just a particular combination of the components of the curvature tensor, 
$G_{\hat{0}\hat{0}}+\sum_{i=1}^{3}G_{\hat{i}\hat{i}}$, 
rather than the energy density of some physical matter fields and 
discuss the possibility of the modification of gravity in the vicinity 
of a massive compact object.

\section{Conclusion \label{sec:summary}}

In this paper, we have obtained two expressions 
for the periapsis shift of a quasi-circular orbit 
in full order with respect to
the gravitational field 
in a general static spherically symmetric spacetime: 
one in terms of the gravitational mass and the Einstein tensor and the other
in terms of the orbital angular velocity and the Einstein tensor.
It is a particular combination of 
the components of the Einstein tensor on the orbit that makes the periapsis shift of a quasi-circular orbit deviate from that 
in the vacuum spacetime. 

Applying the present approach 
to the weak-field and diffuse regime,
we have shown that the active gravitational mass density, which is relevant to the strong energy condition, 
is responsible for the anomaly of the periapsis shift from its vacuum value.
We have derived the critical density such that if the active gravitational 
mass density is beyond this value, 
the periapsis shift becomes backward and shown that a forward shift greater than 
the vacuum value implies the violation of the strong energy condition in Einstein gravity.

Using the observational data in the literature, we have obtained the observational constraints on the active gravitational mass density of 
dark component on the orbits of the planets in the Solar System and 
on the orbit of S2 star in the Galactic Centre. We have also discussed the possibility of 
the violation of the strong energy condition.

The present result implies that 
the physical mechanisms of apparently different phenomena, the accelerated
cosmic expansion and the excessively forward periapsis shift, 
are common. It is the active gravitational mass density being negative or the violation of 
the strong energy condition.
The matter field with this feature is called dark energy.
So, we can investigate the existence of dark energy 
by the close observation of the orbits of stars 
around a massive object, independently from the observation of the 
expansion law of the Universe.

A lesson from the present analysis is that 
by the periapsis shift experiment, we can access the 
geometrical properties of the spacetime on the orbit of the star
but not directly the central gravitating object except for the gravitational mass,
although they certainly are valuable information. 
It should also be noted that the effects from the physical matter fields and the modification of gravity cannot be distinguished solely by the periapsis shift observation.
This means that if we want to use the periapsis shift as a probe into the central gravitating object and its surrounding region, 
we need to invoke some additional experiments, such as light bending and photon sphere observation and/or 
working assumptions such as modelling dark component physics and/or 
choosing gravitational theories.

Generalising the present framework to include highly noncircular orbits
is clearly important for a wider application and more quantitative discussion. 
To discuss the deviation smaller than that due to the spin of the massive central object,  
we should generalise the present formalism to stationary
axisymmetric spacetimes (cf. Ref.~\cite{Bini:2005dy}). 
These interesting generalisations are left for future work.

\acknowledgments

The authors are 
grateful to H.~Ishihara, T.~Kobayashi, Y.~Kojima, K.~Nakamura, 
K.~Nakashi, Y.~Nakayama, K.-I.~Nakao, M.~Nozawa,
and C.-M.~Yoo 
for their helpful comments. 
We appreciate APCTP for its hospitality during completion of this work.
This work was supported by JSPS KAKENHI Grants 
Nos. JP19K03876, JP19H01895, and JP20H05853 (TH); 
JP19K14715 and JP22K03611(TI); 
JP19H01900 and JP19H00695 (HS, YT).

\appendix

\section{Periapsis shift in the anti-de Sitter spacetime \label{sec:ADS}}
Maximally symmetric spacetimes in four dimensions are known to be 
Minkowski, de Sitter, and anti-de Sitter spacetimes.
These spacetimes are solutions of the Einstein equations 
with a cosmological constant $\Lambda=0$, $>0$, and $<0$, respectively.
The line element in these spacetimes 
is written in the following form:
\begin{equation}
 ds^{2}=-\left(1-\frac{1}{3}\Lambda r^{2}\right)dt^{2}+\left(1-\frac{1}{3}\Lambda r^{2}\right)^{-1}dr^{2}+r^{2}(d\theta^{2}+\sin^{2}\theta d\phi^{2})
\end{equation}
and, hence, from Eq.~(\ref{eq:m}) we have
\begin{equation}
 m=\frac{1}{6}\Lambda r^{3}.
\end{equation}
The nonvanishing tetrad components of the Einstein tensor in these 
spacetimes 
are given by 
\begin{equation}
 G_{\hat{0}\hat{0}}=\Lambda,\quad G_{\hat{1}\hat{1}}=G_{\hat{2}\hat{2}}=G_{\hat{3}\hat{3}}=-\Lambda.
\end{equation}
Since the condition~(\ref{eq:E2L2condition}) 
imposes $\Lambda$ to be negative,
only the anti-de Sitter spacetime is admitted for the existence of a circular orbit.
Note that the active gravitational energy $\rho_{A}$ is positive in this case.
From Eqs.~(\ref{eq:Am0}) and (\ref{eq:Am1}), we obtain $A=4$ and, hence, 
$
\Delta\phi_{p}=-\pi. 
$
That is, the periapses repeatedly appear in every half a round.
 
This can be understood as follows. 
The effective potential $V(r)$ in this case is given by 
\begin{equation}
V(r)=\frac{1-E^{2}}{2}-\frac{\Lambda L^{2}}{6}-\frac{1}{6}\Lambda r^{2}+\frac{1}{2}\frac{L^{2}}{r^{2}},
\end{equation}
which has an extremum if and only if $\Lambda<0$. 
With $\Lambda<0$, this agrees with the effective 
potential of a particle in the isotropic harmonic oscillator potential in classical
Newtonian mechanics up to irrelevant constant terms.
In this potential problem, it is well known that the general orbit is 
an ellipse whose centre is at $r=0$, where 
periapses repeatedly appear 
in every half a round. 

\section{Post-Newtonian and diffuse regime \label{sec:PNNMO}}
If both the expansions in powers of $v^{2}$ and $\alpha$ apply, 
we can expand $\Delta \phi_{p}$ as follows:
\begin{eqnarray}
 \Delta \phi_{p}&=&3\pi \left(\frac{2m}{r}-\frac{\epsilon}{\bar{\epsilon}}\right)
+\frac{27\pi m^{2}}{r^{2}}
-3\pi \left(\frac{8m}{r}\frac{\epsilon}{\bar{\epsilon}}
+\frac{p_{r}+2p_{t}}{\bar{\epsilon}}\right)
+\frac{27\pi}{4}\left(\frac{\epsilon}{\bar{\epsilon}}\right)^{2} \nonumber \\
&& +O\left(v^{6}\alpha^{0}, v^{4}\alpha^{1} , v^{2}\alpha^{2}, v^{0}\alpha ^{3}\right) 
\label{eq:Deltaphip_PNRF_1}
\\
&=&3\pi \left(2r^{2}\omega_{\phi}^{2}-\frac{4\pi }{3}\frac{\epsilon}{\omega_{\phi}^{2}}\right) 
+9\pi r^{4}\omega_{\phi}^{4}
-4\pi^{2}\left(11 r^{2}\omega_{\phi}^{2}\frac{\epsilon}{\omega_{\phi}^{2}}+
\frac{p_{r}+2p_{t}}{\omega_{\phi}^{2}}\right) \nonumber \\
&&+12 \pi ^{3}\left(\frac{\epsilon}{\omega_{\phi}^{2}}\right)^{2}
+O\left(v^{6}\alpha^{0}, v^{4}\alpha^{1} , v^{2}\alpha^{2}, v^{0}\alpha ^{3}\right),
\label{eq:Deltaphip_PNRF_2}
\end{eqnarray}
where the first terms include those of $O(v^{2}\alpha^{0})$ and $O(v^{0} \alpha^{1})$, 
while the second, third, and fourth terms are of $O(v^{4}\alpha^{0})$, 
$O(v^{2}\alpha^{1} )$, and $O(v^{0}\alpha^{2})$, respectively.
In this regime, the periapsis shift in the lowest orders is forward (backward) if 
${2m}/{r}>(<){\epsilon}/{\bar{\epsilon}}$, i.e., $ \epsilon<(>)\epsilon_{c}$, 
where $\epsilon_{c}=\rho_{c}c^{2}$ with $\rho_{c}$ defined by Eq.~(\ref{eq:epsilon_c}). 
The lowest-order terms in the above are not completely identical but 
closely related to the expression
obtained in Ref.~\cite{Khriplovich:2006sq}.

\section{Special matter fields \label{sec:special_matter_fields}}

\subsection{Vanishing radial stress}
In the case of $p_{r}=0$, we can eliminate $p_{t}$ 
from Eq.~(\ref{eq:Am1}) using Eq.~(\ref{eq:contracted_Bianchi}) or 
\begin{equation}
 e^{\lambda}\frac{m}{r^{2}}\epsilon-\frac{2}{r}p_{t}=0.
\end{equation}
The resultant {\it exact} expressions for $A$ are very much simplified 
as follows:
\begin{equation}
 A=1-\frac{6m}{r}+3\frac{\epsilon}{\bar{\epsilon}}
=\frac{1-3r^{2}\omega_{\phi}^{2}}{1+3r^{2}\omega_{\phi}^{2}}+4\pi (1+3r^{2}\omega_{\phi}^{2})\frac{\epsilon}{\omega_{\phi}^{2}}.
\label{eq:A_Sigma=0}
\end{equation}
This reproduces the expression for the Einstein cluster~\cite{Igata:2022rcm}.
The periapsis shift is forward (backward) if $\epsilon<(>)\epsilon_{c}$,
while it vanishes if the equality holds, where $\epsilon_{c}=\rho_{c}c^{2}$ with $\rho_{c}$ defined by Eq.~(\ref{eq:epsilon_c}). The periapsis shift greater than that 
in the vacuum spacetime implies 
the violation of the weak energy condition.
It is interesting that these statements hold in full order with respect to the gravitational field in this case.

\subsection{Perfect fluid}

If $p_{r}=p_{t}=p$, for which the matter field is called a perfect fluid, we find 
\begin{eqnarray}
 A&=&\left(1-\frac{2m}{r}\right)\frac{\bar{\epsilon}+3\epsilon+12 p }
{\bar{\epsilon}+3 p }
-\frac{4m}{r}\frac{4\bar{\epsilon}-3\epsilon+9 p }{4\bar{\epsilon}} \\
&=&
\frac{1-3r^{2}\omega_{\phi}^{2}}{1+3r^{2}\omega_{\phi}^{2}}
+\frac{4\pi}{\omega_{\phi}^{2}}
\left[
\left(1+2r^{2}\omega_{\phi}^{2}\right)\epsilon+\frac{3(1+5r^{2}\omega_{\phi}^{2}+2r^{4}\omega_{\phi}^{4})}{1+3r^{2}\omega_{\phi}^{2}} p 
\right].
\end{eqnarray}
The series expansions in different regimes can be obtained by just putting $p_{r}=p_{t}= p $
in the corresponding expressions for general dark component.
Note that the expression in the weak-field and diffuse regime
is consistent with that in Ref.~\cite{Iwata:2016ivt}.

\end{document}